# The Untold Impact of Learning Approaches on Software Fault-Proneness Predictions

Mohammad Jamil Ahmad[a,1], Katerina Goseva-Popstojanova[a,*] and Robyn R. Lutz[b]

[a]*Lane Department of Computer Science and Electrical Engineering, West Virginia University, Morgantown, WV, 26505, USA*
[b]*Department of Computer Science, Iowa State University, Ames, Iowa, 50011, USA*

**ABSTRACT**

Software fault-proneness prediction is an active research area, with many factors affecting prediction performance extensively studied. However, the impact of the learning approach (i.e., the specifics of the data used for training and the target variable being predicted) on the prediction performance has not been studied, except for one initial work. This paper explores the effects of two learning approaches, *useAllPredictAll* and *usePrePredictPost*, on the performance of software fault-proneness prediction, both within-release and across-releases. The empirical results are based on data extracted from 64 releases of twelve open-source projects. Results show that the learning approach has a substantial, and typically unacknowledged, impact on the classification performance. Specifically, using *useAllPredictAll* leads to significantly better performance than using *usePrePredictPost* learning approach, both with-in-release and across-releases. Furthermore, this paper uncovers that, for within-release predictions, this difference in classification performance is due to different levels of class imbalance in the two learning approaches. When class imbalance is addressed, the performance difference between the learning approaches is eliminated. Our findings imply that the learning approach should always be explicitly identified and its impact on software fault-proneness prediction considered. The paper concludes with a discussion of potential consequences of our results for both research and practice.

## 1. Introduction

The prediction of *fault-prone* software units helps software developers prioritize their efforts, reduces development costs, and leads to better quality software products (Nagappan et al., 2006, 2008). It is thus not surprising that the prediction of software fault-proneness is an active research area in software engineering.

A software bug (i.e., fault) is an accidental condition which, if encountered, may cause the software system or component to fail to perform as required (Hamill and Goseva-Popstojanova, 2009). Software bugs can lead to different types of failures, some with serious consequences such as private information leakage, financial loss, or loss of human life. A software unit (e.g., file, package, or component) is *fault-prone* if it has one or more software bugs. Note that in addition to 'fault-prone', the terms 'bug-prone', 'error-prone', and 'defective' have been used.

Over the years, researchers have built many *software fault-proneness prediction models* (Arisholm et al., 2010; Catal, 2011; Hall et al., 2012; Song et al., 2019; Hosseini

*Corresponding author

Mohammad.ahmad@mail.wvu.edu (M.J. Ahmad);
Katerina.Goseva@mail.wvu.edu (K. Goseva-Popstojanova);
rlutz@iastate.edu (R.R. Lutz)

https://engineering.wvutech.edu/faculty-and-staff/mohammad-ahmad-ph-d
(M.J. Ahmad); https://community.wvu.edu/~kagoseva/ (K.
Goseva-Popstojanova); https://robynlutz.com (R.R. Lutz)
ORCID(s): 0000-0002-4038-2379 (M.J. Ahmad); 0000-0003-4683-672X
(K. Goseva-Popstojanova); 0000-0001-5390-7982 (R.R. Lutz)

[1]Present affiliation: Department of Computer Science and Information Systems, West Virginia University Institute of Technology, Beckley, WV, 25801, USA

et al., 2019). These models either classify software units as fault-prone and not fault-prone (e.g., (Koru and Liu, 2005; Nagappan et al., 2006; Menzies et al., 2007; Zimmermann et al., 2007, 2009; Krishnan et al., 2013; Alshehri et al., 2018; Gong et al., 2021)) or predict the number of faults (i.e., fault-count) in each software unit (e.g., (Zimmermann et al., 2007; Ostrand et al., 2005; Devine et al., 2012, 2016)). This paper focuses on the classification-based predictions.

Many papers have explored different factors that affect the performance of the classification-based software fault-proneness prediction. These factors include: choice of machine learning algorithms *(i.e., learners)* (Lessmann et al., 2008; Krishnan et al., 2013), choice of software metrics (i.e., features used for prediction) (Bluemke and Stepień, 2016; Wang et al., 2016; Alshehri et al., 2018; Gong et al., 2021), effect of data balancing techniques (Khoshgoftaar et al., 2010; Wang and Yao, 2013; Song et al., 2019; Goseva-Popstojanova et al., 2019), and datasets used in building prediction models (Goseva-Popstojanova et al., 2019).

However, the specifics of the data used for training and the target variable that the models were designed to predict often were omitted or merely implied by the datasets being used, rather than being explicitly described. In this paper, we use the term *learning approach* to refer to the specifics of the data used for training and to the target (i.e., response) variable, and we explore the impact of the learning approach on the classification performance of software fault-proneness prediction.

Specifically, we consider two learning approaches, *useAllPredictAll* and *usePrePredictPost*, as most software fault-proneness prediction papers can be categorized as using one of them. Both learning approaches use static code metrics collected at a particular time, typically at the release date. However, as shown in Table 1, the learning approaches use different durations for extracting longitudinal metrics, such as change metrics and socio-technical metrics, and different target variables.

**Table 1**
Details on features' collection and target (i.e., response) variable for each learning approach

| | | useAllPredictAll | usePrePredictPost |
|---|---|---|---|
| Features | Static code metrics (snapshot) | collected at a particular time (typically at release date) | collected at a particular time (typically at release date) |
| | Longitudinal metrics such as change metrics, socio-technical metrics | collected for the entire release duration | collected for the pre-release duration |
| | Fault-proneness as input metric | none | pre-release bugs |
| Target variable | Fault-proneness response metric | bugs during the entire release duration (no distinction between pre-release and post-release bugs) | post-release bugs |

Thus, software fault-proneness prediction models that follow the *useAllPredictAll* learning approach use longitudinal features collected from the entire period of a release/project duration for training in order to predict all bugs during the same period, without distinguishing between pre-release and post-release bugs. On the other hand, prediction models that follow the *usePrePredictPost* learning approach use longitudinal features collected during pre-release (including pre-release bugs) in order to predict post-release bugs (i.e., bugs detected after the release date).

Prior works, apart from Krishnan et al. (2013), have not discussed the distinctions between learning approaches when building software fault-proneness prediction models nor explored the impact of the learning approach on the prediction performance.

In addition to the learning approach, our analysis includes the *prediction style* that captures what data were used for training and testing of the prediction models. Software fault-proneness prediction models that are trained and tested on data collected from the same software project are known as *within-project* prediction models. If a software project has multiple releases, the fault-proneness prediction can be done either *within-release* or *across-releases*. Prediction within-release uses training and testing data from the same release, while across-releases prediction uses training data from one or more releases and testing data from a different release. Models that use training data from one project and testing data from a separate project are known as *across-projects* prediction models. Some researchers also use the term across-company prediction for cases where prediction models are trained on data from a project developed by one company and tested on data collected from a project developed by a different company.

The goal of this paper is to study the impact of the learning approach on the software fault-proneness prediction for within-project prediction styles (i.e., within-release and across-releases). Thus, we address the following research questions:

**RQ1:** Does the learning approach affect the classification performance of the software fault-proneness predictions?

**RQ2:** For a given learning approach, what is the difference in classification performance between using within-release and across-releases prediction styles?

The empirical results presented in this paper are based on datasets we extracted from 64 releases of 12 Apache open-source projects.

The main contributions of this study are as follows:

- We categorize the related works on software fault-proneness prediction based on the learning approach used in each work. To the best of our knowledge such categorization has not been done previously. Our identification of the learning approach(es) used in each paper enables a fuller understanding of the trends and results reported by many studies.

- We explore the effects of a hidden variable, learning approach, on the performance of software fault-proneness prediction models. For that purpose, we: (1) build software fault-proneness prediction models using the two learning approaches, *useAllPredictAll* and *usePrePredictPost*; (2) predict fault-proneness both within-release and across-releases; and (3) use a design of experiment (DoE) approach and statistical analysis to draw sound conclusions.

- We present results showing that the learning approach significantly affects the classification performance. Specifically, using the *useAllPredictAll* learning approach led to significantly higher Recall, Precision, F-Score, G-Score and similar 1 - False Positive Rate (1 - FPR) compared to using the *usePrePredictPost* learning approach.

- In investigating why there is a difference in classification performance when using different learning approaches, we uncover that for within-release predictions the difference is due to another hidden variable – the class imbalance. The treatment of imbalance (here using SMOTE) improves the within-release prediction performance for each learning approach individually, as expected based on prior works.

Extending the prior works, we apply class imbalance treatment on both learning approaches and compare the corresponding improvements. Furthermore, based on descriptive and inferential statistical results, we find that the imbalance treatment eliminates the difference in prediction performance between the learning approaches, for within-release prediction.

- We discuss the implications of our findings and provide recommendations for designing, reporting, and comparing software fault-proneness prediction studies, as well as identify some open issues for future research in this area.

In brief, this paper provides evidence that the learning approach must be explicitly considered as an additional factor in the analysis of the software fault-proneness prediction performance. We show that this is due to the fact that the learning approach is intrinsic both to the way datasets are created and to the way machine learning models are trained and tested.

The remainder of the paper is organized as follows. Section 2 describes the related works, and Section 3 categorizes the related works based on the learning approach each used. Section 4 describes the datasets, data extraction process, and feature vectors used in this study. Section 5 details our machine learning approach and design of experiments approach. Analysis of our results are provided in Section 6, with threats to validity presented in Section 7. Section 8 discusses the implications of our findings and offers associated recommendations toward improved fault-proneness prediction performance. Section 9 provides concluding remarks.

## 2. Related work

Software fault-proneness prediction is an active area of research, as evidenced by its systematic literature review papers (Arisholm et al., 2010; Catal, 2011; Hall et al., 2012; Song et al., 2019; Hosseini et al., 2019) and the references therein. We here discuss the related studies that are the most relevant to our work.

Many different machine learning algorithms have been used in building software fault-proneness prediction models. These include J48 (Moser et al., 2008; Kamei et al., 2010; Krishnan et al., 2013), Random Forest (RF) (Guo et al., 2004; Mahmood et al., 2018; Fiore et al., 2021; Gong et al., 2021), and combinations of several machine learning algorithms, e.g., OneR, J48, and Naïve Bayes (NB) in (Menzies et al., 2007), RF, NB, RPart, and SVM in (Bowes et al., 2018), J48, RF, NB, Logistic Regression (LR), PART, and G-Lasso in (Goseva-Popstojanova et al., 2019), and Decision Tree (DT), k-Nearest Neighbor (kNN), LR, NB, and RF in (Kabir et al., 2021). With recent advances in Deep Neural Networks (DNN), some software fault-proneness prediction studies used deep learning (Wang et al., 2016; Li et al., 2017; Pang et al., 2017; Zhou et al., 2019; Zhao et al., 2021).

Software fault-proneness prediction models utilize feature vectors consisting of software metrics extracted from software source or binary code, its development history, and the associated bug tracking systems. In general, the extracted software metrics can be static code metrics, change metrics, or social metrics. Static code metrics are collected from the software source code or binary code units (Koru and Liu, 2005; Menzies et al., 2007; Lessmann et al., 2008; Menzies et al., 2010; He et al., 2013; Ghotra et al., 2015; Bowes et al., 2018; Kabir et al., 2021). Change metrics, sometimes called process metrics, are collected from the projects' development history (i.e., commit logs) and bug tracking systems (Nagappan et al., 2010; Giger et al., 2011; Krishnan et al., 2011, 2013; Goseva-Popstojanova et al., 2019). Social metrics are extracted from the communications among developers and/or users of a software project (Bird et al., 2009). Some studies used only static code metrics (Menzies et al., 2007; Mende and Koschke, 2009; Song et al., 2011; Xu et al., 2018, 2019; Amasaki, 2020; Kabir et al., 2021), only change metrics (Giger et al., 2011; Krishnan et al., 2011, 2013; Goseva-Popstojanova et al., 2019), organizational metrics (Nagappan et al., 2008) or metrics derived from the contribution networks (Pinzger et al., 2008). There are studies that used combinations of different types of metrics (Nagappan et al., 2010; Arisholm et al., 2010; Giger et al., 2012; Alshehri et al., 2018), including studies that combined metrics extracted from contribution networks, dependency networks, and/or socio-technical networks (Bird et al., 2009; Gong et al., 2021).

Works that do prediction at the file level use features collected from units such as files (Zimmermann et al., 2007; Moser et al., 2008; Kamei et al., 2010; Krishnan et al., 2013; Goseva-Popstojanova et al., 2019), classes (Koru and Liu, 2005; Malhotra and Raje, 2015; Song et al., 2019), or methods (Giger et al., 2012; Bowes et al., 2018), while prediction at the component level is based on features aggregated at the package, module, or component level from features extracted at file, class, or method level (Zimmermann et al., 2007).

Next, we summarize related works by the prediction style they used. Note that 'prediction style' has elsewhere been referred to as 'context' (Gong et al., 2021). For within-project predictions, some works were focused only on within-release prediction style (e.g., (Nagappan et al., 2008; Krishnan et al., 2011, 2013; Alshehri et al., 2018)), while others focused only on across-releases prediction (Xu et al., 2018, 2019; Fiore et al., 2021; Kabir et al., 2021). Terms used in the literature for 'across-releases' prediction include 'cross-version' (Xu et al., 2018, 2019; Fiore et al., 2021; Gong et al., 2021) and 'inter-release' prediction (Kabir et al., 2021). Papers that included across-projects prediction models include (Nagappan et al., 2006; Zimmermann et al., 2009; He et al., 2013; Wang et al., 2016; Gong et al., 2021). Other papers specifically explored the across-company prediction style (Turhan et al., 2009). In this paper we focus on the two within-project predictions styles – within-release and across-releases – and in Section 3 categorize related works by the learning approach and prediction style used (see Table 2).

Many software fault-proneness prediction studies have derived conclusions solely from analyzing the empirical results, without applying statistical tests (Bird et al., 2009; Giger et al., 2012; Alshehri et al., 2018; Zhou et al., 2019; Fiore et al., 2021). Other studies used statistical inference to support their conclusions, typically studying one factor at a time (e.g., (Mende and Koschke, 2009; Turhan et al., 2009; Song et al., 2011; Krishnan et al., 2013; Okutan and Yıldız, 2014; Bowes et al., 2018; Gong et al., 2021)).

The design of experiment approach (DoE), which we use in our investigation here, allows one or more input factors to be systematically explored with the goal of determining their effect on the output (response) variable. DoE has been used to date only in several software fault-proneness prediction studies (Khoshgoftaar and Seliya, 2004; Lessmann et al., 2008; Gao et al., 2011; Shepperd et al., 2014; Tantithamthavorn et al., 2016; Shepperd et al., 2018)).

Khoshgoftaar and Seliya (2004) treated the machine learning algorithm as a factor and the software release as a block. The results, based on four successive releases from a large legacy telecommunication system, showed that the models' predictive performances were significantly different across the releases, implying that the predictions were influenced by the characteristics of the data. Lessmann et al. (2008) explored the effect of the machine learning algorithm on the software fault-proneness prediction performance using ten datasets from the NASA Metrics Data Program (MDP) repository. The results showed that no statistically significant difference in performance existed among the top 17 (out of 22) machine learning algorithms, i.e., the choice of machine learning algorithm appeared to be not as important as was previously assumed. Gao et al. (2011) statistically examined the effect of three factors (i.e., feature ranking method, feature subset selection method, and machine learning algorithm) on the software fault-proneness prediction performance for a large legacy telecommunication software. Shepperd et al. (2014) presented meta-analysis based on 42 primary software fault-proneness prediction studies. Note that substantial overlaps existed in dataset usage among the primary studies, with the NASA datasets family dominating with 59%, followed by the Eclipse family with 21%. This work used DoE consisting of four factors: the machine learning algorithm, dataset used, input metrics (i.e., features), and the particular researcher group (i.e., the authors). The results showed that the researcher group contributed the most to the variance of the prediction performance, followed by the dataset, and input metrics. The choice of the machine learning algorithm had the least impact on the prediction performance. Two follow-up studies (Tantithamthavorn et al., 2016; Shepperd et al., 2018) repeated the meta-analysis using subsets of the primary studies used in (Shepperd et al., 2014). The work in (Tantithamthavorn et al., 2016), which was based on the Eclipse datasets family, found that the research group had smaller impact than the input metrics. When only the NASA datasets family was used for re-analysis in (Shepperd et al., 2018), the results were more in line with the initial analysis based on all datasets families (Shepperd et al., 2014).

In summary, prior DoE studies had different goals than our study and looked at different factors. None of these studies recognized the learning approach(es) used and some mixed different learning approaches without explicit acknowledgement (Shepperd et al., 2014).

In this paper we will show how the choice of the learning approach significantly affects the prediction performance and therefore must be explicitly specified and accounted for in the analysis. Additionally, we will show how awareness of the learning approach factor provides improved understanding of the impact of this factor on the prediction results. Thus, researchers and practitioners alike must recognize that different datasets are not only extracted from artifacts of different software systems, but also may be produced to fit different learning approaches. We believe that considering the learning approach explicitly as a factor will help disentangle the effects of the other factors, since the learning approach is intrinsic both to the datasets used and the learning process conducted.

## 3. Categorizing related works by learning approach used

To better understand how the selection of learning approach influences fault-proneness prediction results, we need to first identify the learning approach that was used in existing studies. However, when we set out to categorize the learning approach used in each of those studies, we found that this was a difficult task. In fact, as we will see in this section, apart from Krishnan et al. (2013), studies on software fault-proneness prediction have not explicitly specified the learning approach nor described the impact of the learning approach used on the prediction performance they reported. It appears that the characteristics of the dataset(s) used for building the software fault-proneness prediction models typically predetermined which learning approach was used. By carefully exploring the details of the datasets being used, we were able to categorize the related works according to the learning approach.

The results of our effort are shown in Table 2, which groups the related studies by: the dataset(s) used, the learning approach (i.e., *useAllPredictAll* or *usePrePredictPost*), and the prediction style (i.e., within-release or across-releases). Those studies which used prediction within-project with no releases (e.g., studies based on the NASA MDP datasets) are grouped with the studies which used prediction within-release because they are designed similarly. Studies which used multiple datasets attributed to one specific learning approach (e.g., (Kim et al., 2011; Bowes et al., 2018)) or to multiple learning approaches (e.g., (Nam et al., 2013; Shepperd et al., 2014; Song et al., 2019; Zhou et al., 2019)) are shown in more than one cell.

Next, we discuss the related works that employed only *useAllPredictAll* learning approach (subsection 3.1), only *usePrePredictPost* learning approach (subsection 3.2), as well as those related works that used datasets or results of studies that belong to both learning approaches (subsection 3.3).

**Table 2**
Related work studies categorized by the learning approach and prediction style

| Dataset | useAllPredictAll | | usePrePredictPost | |
|---|---|---|---|---|
| | within-release/ project | across-releases | within-release/ project | across-releases |
| NASA MDP | Koru and Liu (2005); Menzies et al. (2007); Gondra (2008); Lessmann et al. (2008); Jiang et al. (2008a,b,c); Elish and Elish (2008); Turhan et al. (2009); Mende and Koschke (2009); Menzies et al. (2010); Wang and Yao (2013); Shepperd et al. (2014); Ghotra et al. (2015); Bowes et al. (2018); Shepperd et al. (2018); Zhou et al. (2019); Goyal (2022) | | | |
| Apache, other open source and industry projects (Jureczko and Madeyski, 2010; Jureczko and Spinellis, 2010) | He et al. (2013); Okutan and Yıldız (2014); Bowes et al. (2018); Song et al. (2019) | Madeyski and Jureczko (2015); He et al. (2015); Li et al. (2017); Xu et al. (2018, 2019); Amasaki (2020); Kabir et al. (2021) | | |
| Eclipse plugins | Giger et al. (2011, 2012) | | | |
| Eclipse platform | Kim et al. (2011); Krishnan et al. (2013); Shepperd et al. (2014) | | Zimmermann et al. (2007); Moser et al. (2008); Bird et al. (2009); Kamei et al. (2010); Krishnan et al. (2013); Shepperd et al. (2014); Tantithamthavorn et al. (2016); Alshehri et al. (2018); Shepperd et al. (2018); Goseva-Popstojanova et al. (2019) | Zimmermann et al. (2007); Bird et al. (2009); Kamei et al. (2010); Shepperd et al. (2014); Tantithamthavorn et al. (2016) |
| Open-source | Shepperd et al. (2014); Malhotra and Raje (2015); Ghotra et al. (2015) | Malhotra and Raje (2015); Fiore et al. (2021) | Shepperd et al. (2014); Goseva-Popstojanova et al. (2019); Gong et al. (2021) | Wang et al. (2016); Gong et al. (2021) |
| Microsoft products | Layman et al. (2008) | | Nagappan et al. (2008); Bird et al. (2009); Nagappan et al. (2010); Shepperd et al. (2014) | Nagappan et al. (2006) |
| Telecommunication software | Tosun et al. (2010); Shepperd et al. (2014); Bowes et al. (2018) | Arisholm et al. (2007); Shepperd et al. (2014) | Arisholm et al. (2010); Shepperd et al. (2014) | Khoshgoftaar and Seliya (2004); Arisholm et al. (2010); Gao et al. (2011); Shepperd et al. (2014) |
| AEEEM (D'Ambros et al., 2010) | | | Nam et al. (2013); Song et al. (2019); Zhou et al. (2019) | |
| Relink (Wu et al., 2011) | Wu et al. (2011); Nam et al. (2013); Zhou et al. (2019) | | | |
| Other | Turhan et al. (2009); Menzies et al. (2010); Kim et al. (2011); Giger et al. (2012); Wang and Yao (2013); Shepperd et al. (2014) | Madeyski and Jureczko (2015) | Shepperd et al. (2014) | |

### 3.1. *useAllPredictAll* learning approach

Software fault-proneness prediction models employing *useAllPredictAll* learning approach are trained on data collected from the entire duration of each release or project with the goal of predicting fault-prone files within the same period. These models do not distinguish between pre-release and post-release bugs. That is, all bugs are grouped in the *bug-fixes* metric, and a file is labeled as fault-prone if it had a bug anytime during the specific release or project.

Based on our analysis of the related work, studies that used the NASA MDP datasets (e.g., those available on PROMISE (Sayyad and Menzies, 2005)) have followed the *useAllPredictAll* learning approach. Note that NASA datasets have no releases; each represents an independent project for which static code metrics were extracted as snapshots of source code at a given time. Examples of studies which used the NASA MDP datasets within-project include but are not limited to (Koru and Liu, 2005; Menzies et al., 2007; Gondra, 2008; Lessmann et al., 2008; Jiang et al., 2008a,b,c; Elish and Elish, 2008; Turhan et al., 2009; Mende and Koschke, 2009; Menzies et al., 2010; Wang and Yao, 2013; Ghotra et al., 2015; Bowes et al., 2018; Zhou et al., 2019; Goyal, 2022).

Other datasets that utilized the *useAllPredictAll* learning approach were created from fifteen open-source projects (including twelve Apache projects) with a total of 48 releases and six industrial projects with a total of 27 releases (Jureczko and Spinellis, 2010; Jureczko and Madeyski, 2010) and were donated to PROMISE (Sayyad and Menzies, 2005). These datasets have a set of static code metrics as feature vectors and were used by multiple studies for prediction both within-release (He et al., 2013; Okutan and Yıldız, 2014; Bowes et al., 2018) and across-releases (Madeyski and Jureczko, 2015; He et al., 2015; Li et al., 2017; Xu et al., 2018, 2019; Amasaki, 2020; Kabir et al., 2021).

Different datasets obtained from Eclipse were extensively used for software fault-proneness prediction. Datasets suitable for *useAllPredictAll* learning approach were employed to predict software fault-proneness of Eclipse plugins (Giger et al., 2011, 2012), as well as of Eclipse platform (Kim et al., 2011; Krishnan et al., 2013). These four works used change metrics as feature vectors.

Other open-source projects were used to extract datasets adequate for the *useAllPredictAll* learning approach, including Android (Malhotra and Raje, 2015). Some models were used for only within-release predictions (Ghotra et al., 2015) or only across-releases prediction (Fiore et al., 2021), while others were used for both within-release and across-releases predictions (Malhotra and Raje, 2015).

The *useAllPredictAll* learning approach was also used for software fault-proneness prediction of commercial software, such as within-release predictions for Microsoft products (Layman et al., 2008), and within-release (Tosun et al., 2010; Bowes et al., 2018) and across-releases (Arisholm et al., 2007) for telecommunication software.

Many works used datasets from multiple sources. Examples include NASA MDP, open-source, and telecommunication datasets (Bowes et al., 2018); Eclipse plugins with other open-source projects (Giger et al., 2012); NASA MDP and SOFTLAB telecommunication datasets (Turhan et al., 2009); NASA MDP and datasets from a Turkish manufacturer (Menzies et al., 2010); and datasets obtained from Eclipse platform and other open-source programs (Kim et al., 2011).

### 3.2. *usePrePredictPost* learning approach

Software fault-proneness prediction models which use the *usePrePredictPost* learning approach are trained on features (including pre-release bugs) collected during the pre-release duration to predict post-release bugs. Pre-release and post-release bugs are distinguished, and a file is labeled as fault-prone if it had a post-release bug. This learning approach was first used for software fault-proneness prediction of three Eclipse platform releases, within and across-releases, at both file and package level (Zimmermann et al., 2007). The datasets consisted of static code metrics collected from each release. The pre-release bugs were collected six months before the release date (i.e., during the development and testing phase), and the post-release bugs were collected six months after the release date (i.e., after deploying the release to users).

Eclipse datasets created for the *usePrePredictPost* learning approach were used extensively for within-release (Zimmermann et al., 2007; Moser et al., 2008; Bird et al., 2009; Kamei et al., 2010; Krishnan et al., 2013; Tantithamthavorn et al., 2016; Alshehri et al., 2018; Goseva-Popstojanova et al., 2019) and across-releases prediction (Zimmermann et al., 2007; Bird et al., 2009; Kamei et al., 2010). Some of these studies used both static code metrics and change metrics (Moser et al., 2008; Kamei et al., 2010; Alshehri et al., 2018), only static code metrics (Zimmermann et al., 2007), or only change metrics (Krishnan et al., 2013; Goseva-Popstojanova et al., 2019).

Datasets from other open-source projects were also used, sometimes in combination with one or more other datasets (e.g., Eclipse platform, Microsoft products), for within-release predictions (Goseva-Popstojanova et al., 2019), across-releases predictions (Wang et al., 2016), or both (Gong et al., 2021). The *usePrePredictPost* learning approach has also been used for software fault-proneness prediction of commercial software, including Microsoft products for within-release (Nagappan et al., 2008, 2010; Bird et al., 2009) and across-release predictions (Nagappan et al., 2006), as well as for telecommunication software for across-releases predictions (Khoshgoftaar and Seliya, 2004; Gao et al., 2011) and for both within-release and across-releases predictions (Arisholm et al., 2010).

### 3.3. Studies that used datasets from both learning approaches

Some related works used datasets corresponding to different learning approaches for synthesizing the findings across software fault-proneness prediction studies (Hall et al., 2012), to conduct meta-analysis (Shepperd et al., 2014, 2018), or to build and compare prediction models (Nam et al., 2013; Song et al., 2019; Zhou et al., 2019).

The systematic literature review on software fault-proneness prediction (Hall et al., 2012) reported the synthesis of results from 19 classification-based studies, some of which used *useAllPredictAll* while others used the *usePrePredictPost* learning approach. Datasets used in these studies were analyzed in (Hall et al., 2012) as a part of context factors and it was concluded that it may be more difficult to build models for some systems than for others.

The meta-analysis presented in (Shepperd et al., 2014) was based on quantitative results extracted from 42 primary software fault-proneness prediction studies which used datasets that belong to different learning approaches. Most prominently, 59% of the primary studies used the NASA datasets and 21% used the Eclipse datasets, which belong to the *useAllPredictAll* and *usePrePredictPost* learning approaches, respectively. (For other datasets and the corresponding learning approaches used by the primary studies included in (Shepperd et al., 2014) see Table 2.) The results based on using four-way ANOVA, attributed most of the variance to the researcher group and the dataset factors (with 31.0% and 11.2%, respectively). As discussed in Section 2, two follow-up studies (Tantithamthavorn et al., 2016; Shepperd et al., 2018) repeated the meta-analysis using subsets of the primary studies used in (Shepperd et al., 2014). The findings presented in (Tantithamthavorn et al., 2016) were based on using only the Eclipse datasets family, while Shepperd et al. (2018) carried on separate analysis for the Eclipse datasets family and NASA datasets family and compared the results with those presented in (Shepperd et al., 2014; Tantithamthavorn et al., 2016).

With a goal to explore transfer learning, prediction models were built in (Nam et al., 2013) using several datasets from ReLink (Wu et al., 2011) and AEEEM (D'Ambros et al., 2010). The prediction done using ReLink followed the *useAllPredictAll* learning approach, while the AEEEM datasets were extracted following the *usePrePredictPost* learning approach. Another work (Song et al., 2019) explored the role of imbalanced learning on software fault-proneness prediction using the datasets created by Jureczko and Madeyski (2010); Jureczko and Spinellis (2010) and available in PROMISE and the AEEEM datasets (D'Ambros et al., 2010), which correspond to the *useAllPredictAll* and *usePrePredictPost* learning approaches, respectively. The work presented in (Zhou et al., 2019) proposed a deep forest model and compared its performance with RF, NB, LR, and SVM, using the NASA MDP, PROMISE, ReLink, and AEEEM datasets. The first three datasets belong to *useAllPredictAll* learning approach, while the forth dataset corresponds to *usePrePredictPost* learning approach.

To the best of our knowledge, none of the related works on software fault-proneness prediction has addressed the effect of the learning approach on the prediction performance, regardless of whether they used only one learning approach (see subsections 3.1 and 3.2) or both learning approaches (i.e., works discussed in this subsection). An exception is the previous work by Krishnan et al. (2013), which investigated whether the classification performance improved as the Eclipse product line evolved through seven releases. In that work, change metrics were used for within-release prediction only, and the software fault-proneness prediction performance was compared using three different learning approaches *useAllPredictAll*, *useAllPredictPost*, and *usePrePredictPost*.

> One of the contributions of this paper is the categorization of related works by learning approach used, given in Table 2 and discussed in this section. Knowing the learning approach used in each study often explains differences among their prediction performances, and enables better understanding of how fault-proneness prediction works.

Motivated by our categorization here of the related works and by the initial results presented in (Krishnan et al., 2013), in the rest of this paper we explore systematically and rigorously the impact of the learning approach used on the performance of software fault-proneness prediction. Specifically, employing both static code metrics and change metrics extracted from 64 releases of 12 open-source projects, we evaluate the performance of prediction models both within-release and across-releases. To quantify and better understand the impact of the learning approach on fault-proneness prediction, we use a design of experiments approach and inferential statistical analysis. We show in this paper that the effect of the learning approach is an essential factor in understanding the prediction results. Toward better fault-proneness prediction models, we aim to encourage awareness and attention to the impact of the learning approaches going forward.

## 4. Data collection and building feature vectors

In this section, we first present the data extraction process used for the two learning approaches, then describe the open source projects used to extract the datasets, followed by a description of the feature vectors used for prediction.

### 4.1. Data extraction

We use Figs. 1(A) and 1(B) to illustrate the data extraction for the *useAllPredictAll* and *usePrePredictPost* learning approaches, respectively. As shown in these figures, the release duration of any given release $n$ is the period between the two dates $d1$ and $d2$, shown by the green and the red lines, respectively. Following the approach introduced by Zimmermann et al. (2007), $d1$ is the middle date between the release $n-1$ release date and the release $n$ release date, while $d2$ is the middle date between the release $n$ release date and the release $n+1$ release date.

#### 4.1.1. Data extraction for useAllPredictAll

As illustrated in Fig. 1(A), for each release we extracted the static code metrics from the latest version of the binaries available on the release date. Following the method used in (Krishnan et al., 2013), for each release we extracted the change metrics from the entire duration of that release (i.e., between the $d1$ and $d2$ dates).

When using the *useAllPredictAll* learning approach, no distinction was made between pre-release and post-release bugs; they both were grouped into one metric called *bug-fix*. In other words, a software file had a bug-fix and consequently was labeled as fault-prone if: (i) it was changed by at least one commit which was used to fix a bug, and (ii) that commit was made during the duration of that release, i.e., the period between $d1$ and $d2$ in Fig. 1(A).

Software fault-proneness prediction models using this learning approach were trained using static code metrics extracted on the release date and change metrics extracted from the entire release duration to predict the response variable, bug-fixes, during the entire release duration.

#### 4.1.2. Data extraction for usePrePredictPost

Following the method illustrated in Fig. 1(B), previously used in (Nagappan et al., 2006; Zimmermann et al., 2007; Moser et al., 2008; Krishnan et al., 2013), for each release we extracted the change metrics from the pre-release duration only, that is, for the period between $d1$ and $n$ in Fig. 1(B). The same static code metrics as for the *useAllPredictAll* learning approach were used (i.e., static code metrics extracted from the latest version of the binaries on the release date).

When using the *usePrePredictPost* learning approach we distinguished the bugs based on when they were detected and fixed. Thus, a software file had a *pre-release bug* if: (i) it was changed by at least one commit to fix a bug, and (ii) that commit was made during the pre-release duration, i.e., the period between $d1$ and $n$ in Fig. 1(B). A software file had a *post-release bug* and consequently was labeled as fault-prone if: (i) it was changed by at least one commit to fix a bug, and (ii) that commit was made after the release date, i.e., the period between $n$ and $d2$ in Fig. 1(B).

Software fault-proneness prediction models using the *usePrePredictPost* learning approach were trained using the static code metrics extracted at the release date, the change metrics extracted pre-release, and the pre-release bugs to predict the response variable post-release bugs.

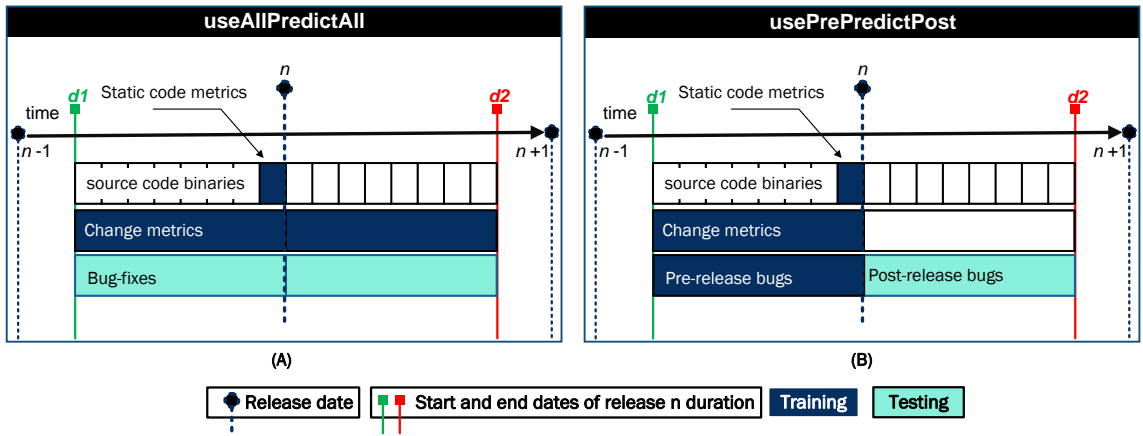

**Figure 1**: Metrics extraction process for the two learning approaches

### 4.2. Open-source projects used to extract datasets

For this work we initially explored 23 projects from the Apache Software Foundation (2022a) that satisfied the reporting criteria defined by (Hall et al., 2012). Some projects, however, did not have clear release dates or a central repository for the source code, or were missing other artifacts. To be included in our study, a project had to satisfy the following six criteria: (i) Availability of the source code or binary distribution, needed to extract code metrics. (ii) Availability of the version control system, needed to generate the commit log file which is used to extract the change metrics. (iii) Availability of the bug tracking system, needed to identify the commits made to fix software bugs. (iv) At least two releases of the project, in order to build predictors across-releases (i.e., training the model on one release to predict fault-proneness for the subsequent release). (v) Release dates clearly specified, in order to be able to extract change metrics for each release as described in subsection 4.1. (vi) Release dates within the commits log file dates, to ensure that all commits made within a given release are considered.

Out of the 23 projects initially considered, eleven did not satisfy one or more of these inclusion criteria and therefore were excluded from this study. The names of these eleven projects and the specific inclusion criteria that were violated are provided in (Ahmad et al., 2022).

The datasets used in this study were extracted from the twelve open-source projects that satisfied all inclusion criteria and are listed in Table 3. Following the reporting requirements given in (Hall et al., 2012), for each project, Table 3 provides the number of releases, size in LOC, number of developers, maturity in years, and application domain/functionality. All projects were written in Java. The classes which are interfaces were excluded from the datasets.

The total number of files, and the percentages of fault-prone files, for all releases of all twelve projects are provided in (Ahmad et al., 2022). For brevity, here we present the box plots of the percentages of fault-prone files: (1) for multiple releases of each of the twelve projects in Fig. 2, and (2) for all releases of all projects cumulatively in Fig. 3.

As shown in these figures, the percentage of fault-prone files is significantly higher for the *useAllPredictAll* learning approach compared to the *usePrePredictPost* learning approach, both for the releases of each project individually (Fig. 2) and for all projects cumulatively (Fig. 3). For example, for the least fault-prone project, MyFaces, the mean percentages of fault-prone files across its four releases were 14.9% for *useAllPredictAll* and only 2.6% for *usePrePredictPost* learning approach. Similarly, the most fault-prone project, jEdit, had 69.5% and 28.3% mean percentages of fault-prone files across its four releases for the *useAllPredictAll* and *usePrePredictPost* learning approaches, respectively. These observations are explained by the facts that *useAllPredictAll* learning approach accounts for fault-prone files during the entire release duration, while the *usePrePredictPost* learning approach only considers post-release fault-prone files.

### 4.3. Feature vectors used for prediction

We extracted 20 static code metrics and 14 change metrics at the file level, which were then combined in feature vectors and used in our models. Note that while static code metrics represent a snapshot in time, change metrics capture the longitudinal changes made to each file over the specified duration (see section 4.1).

**Static code metrics.** We extracted 20 widely used static code metrics at the file level from the binary code of each release, on the release date from the Archive Server of the Apache Software Foundation (2022a). The metrics were extracted using the Chidamber and Kemerer Java Metrics (CKJM) tool (Jureczko and Spinellis, 2011). They belong to six different metric suites: C&K metrics, Henderson-Sellers, Martin, QMOOD, Tang, and McCabe. The full descriptions of the static code metrics can be found in (Jureczko and Madeyski, 2010; Ahmad et al., 2022).

**Change metrics.** We extracted 14 change metrics at the file level following the process described in (Moser et al., 2008; Krishnan et al., 2013). Change metrics were extracted by integrating the information from the version control system (i.e., commit log) and the bug tracking system of each project. All twelve projects included in our study used the SVN version control system. Eleven of these projects used the JIRA bug tracking system (Jira, 2022), and one (i.e., Ant) used the Bugzilla bug tracking system (Apache Software Foundation, 2022b). Change metrics for each release were extracted from all commits made between $d1$ and $d2$ dates for the *useAllPredictAll* learning approach and between $d1$ and $n$ dates for the *usePrePredictPost* learning approach, as shown in Figs. 1(A) and 1(B), respectively. The list of change metrics along with their definitions and additional details about the change metrics extraction process are provided in (Ahmad, 2021; Ahmad et al., 2022).

**Table 3**
List of projects used in this study and their details. All projects were implemented in Java.

| Project | # releases | Size (LOC) | # developers | Maturity (years) | Domain / functionality |
|---|---|---|---|---|---|
| Ant | 7 | 376,250 | 47 | + 10 | Command-line tool for java application building |
| Axis2 | 5 | 409,432 | 29 | + 10 | Web services creating and usage |
| Derby | 9 | 1,759,271 | 34 | + 9 | Relational Database |
| Geronim | 5 | 581,083 | 49 | + 8 | Libraries for JavaEE/JakartaEE |
| Hadoop | 6 | 1,500,351 | 72 | + 7 | Distributed computing platform |
| Hive | 6 | 3,255,810 | 43 | + 7 | Data warehousing |
| jEdit | 4 | 346,197 | 40 | + 15 | Programmer text editor |
| MyFace | 4 | 377,930 | 36 | + 9 | Sub-projects for JavaServer technology |
| Pivot | 5 | 215,406 | 7 | + 6 | Platform for building install-able Internet Applications |
| Synapse | 3 | 376,250 | 23 | + 10 | Web Services |
| Wicket | 6 | 398,043 | 21 | + 8 | Web-apps developing environment |
| Xalan | 4 | 398,183 | 32 | + 14 | XSLT processor |

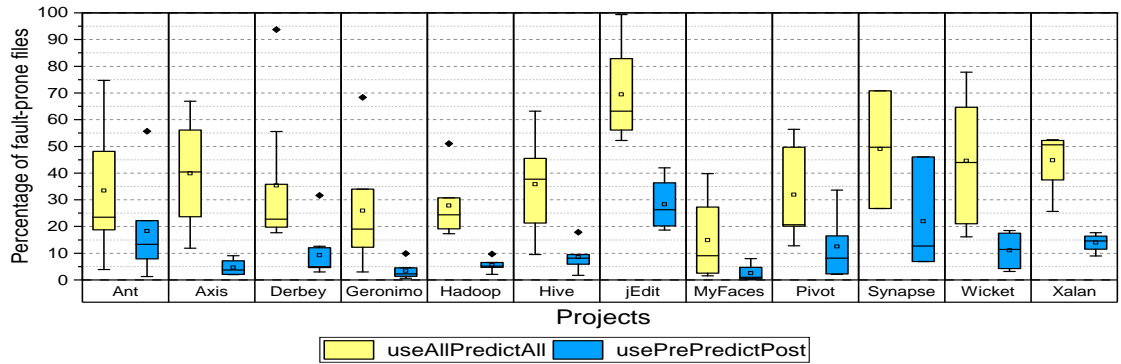

**Figure 2:** Box plots of the percentages of fault-prone files per project, each with multiple releases

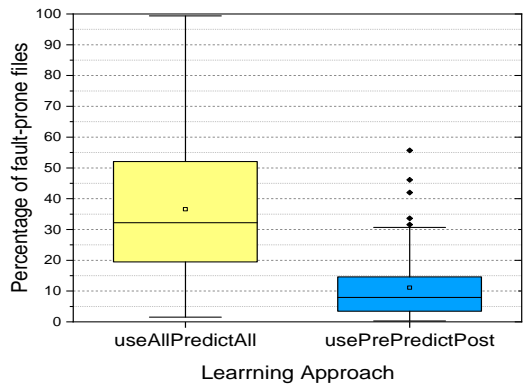

**Figure 3:** Box plots of the percentages of fault-prone files for all releases of all projects

**Feature vectors** used for our machine learning experiments integrate the static code and change metrics which were extracted separately. To combine these metrics for each file, in each release and project, we created an algorithm, provided in (Ahmad et al., 2022), that matches the names of class files from the binary distribution with the names of Java files from the commit logs. Note that the number of files in the static code metrics list was higher than the number of files in the change metrics list because the binary distributions of each release also had other files from external software libraries, which were excluded from the analysis. For the files considered in this work, if the class name from the static code metrics list matched the file name in the commits logs for that release, the changes were aggregated, and the change metrics were created. Otherwise, it meant that the file did not change, and the values of all its change metrics were set to 0 for that release. The full feature vector for each file was created by concatenating the 20 static code and 14 change metrics. As described in subsection 4.1, a file was labeled as fault-prone if it had at least one bug-fix in the case of the *useAllPredictAll* learning approach or at least one post-release bug in the case of the *usePrePredictPost* learning approach.

## 5. Machine learning and DoE approaches

This section presents our machine learning approach, which utilizes the created feature vectors for prediction of fault-proneness, followed by the description of the DoE approach we used to explore and quantify the effect of the learning approach on the prediction performance.

### 5.1. Machine learning approach

For our experiments, we used several well-known machine learning algorithms: Linear Discriminant Analysis (LDA), kNN, LR, NB, J48, and RF. The results showed that, for almost all performance metrics, there were no statistically significant differences between J48, LDA, and LR. RF performed slightly better, while kNN and NB performed worse than these three learners. Since the effect of the learner on prediction performance has been previously explored by many works, we do not treat it as a factor in our experiments. We here only present the findings of using the J48 algorithm, which has been widely used for software fault-proneness prediction (Khoshgoftaar and Seliya, 2004; Guo et al., 2004; Arisholm et al., 2007; Menzies et al., 2007; Moser et al., 2008; Arisholm et al., 2010; Krishnan et al., 2013; Alshehri et al., 2018; Goseva-Popstojanova et al., 2019) and has been shown to be among the top performing algorithms (Lessmann et al., 2008; Krishnan et al., 2013; Shepperd et al., 2014). The findings obtained when using the other learning algorithms were consistent with those presented here, e.g., the results of using RF are provided in the supplemental document online (Ahmad et al., 2022).

We evaluated models' prediction performance using the feature vectors consisting of both the static code metrics and change metrics (see subsection 4.3).

For each learning approach, we explored both within-release and across-releases predictions. For within-release prediction, we built the models using two data splitting techniques: 10-fold cross validation and 50/50 split. In both cases, the models were trained and tested on data collected from the same release. For 10-fold cross validation (which we label as M1) the data were divided into ten folds using random stratified sampling; nine folds of the data were used for training and the remaining fold was used for testing. This was repeated ten times, each time using a different fold for testing. 10-fold cross validation is widely used in this field (Turhan et al., 2009; Giger et al., 2012; Bowes et al., 2018).

For the second data splitting technique, referred to as 50/50 split (and labeled as M2), stratified random sampling was used to split the data into two folds, with 50% of the data in each. One fold was used to train the model and the other to test it. To avoid bias, following (Kamei et al., 2010; Nam et al., 2013), the 50/50 split was repeated 100 times with different random stratified samples, and the averages of the performance metrics are reported. We chose this splitting technique as it is similar in design to the across-release prediction.

For the prediction across-releases (labeled as M3), the entire data from a given release $n$ was used for training, and the entire data from release $n + 1$ was used for testing (Arisholm et al., 2007; Nam et al., 2013; Song et al., 2019). This was repeated 100 times and the average performance metrics were reported.

For evaluating the models' classification performance, we used the following metrics:

Recall (R) = TP/(TP + FN)

Precision (P) = TP/(TP + FP)

False Positive Rate (FPR) = FP/(FP + TN)

F-Score = 2 · Recall · Precision/(Recall + Precision)

G-Score = 2 · Recall · (1-FPR)/ [Recall + (1-FPR)]

where True Positive (TP) represents the number of files that were faulty, and were predicted to be faulty; False Negative (FN) represents the number of files that were faulty, but were predicted to be not faulty; False Positive (FP) represents the number of files that were not faulty, but were predicted to be faulty; and True Negative (TN) represents the number of files that were not faulty and were predicted to be not faulty.

The values of all performance metrics are between 0 and 1. Good performance has higher Recall and Precision, and lower FPR. F-Score and G-Score are harmonic means of two metrics (i.e., Recall and Precision for F-Score and Recall and (1-FPR) for G-Score) and are high only when both metrics are high. Note that instead of FPR we report (1-FPR), which leads to higher values indicating better performance for all performance metrics.

### 5.2. Design of Experiments (DoE) approach

In this study we used Design of Experiment (DoE) approach to draw statistically sound findings related to the effect of factors (independent variables) on the response variables (i.e., performance metrics of the software fault-proneness prediction). Basically, we considered two factors:

**Factor A:** the learning approach, with two levels:
- *useAllPredictAll*
- *usePrePredictPost*

**Factor B:** the prediction style, with three levels:
- within-release using 10-fold cross validation (**M1**)
- within-release using 50/50 split (**M2**)
- across-releases (**M3**).

In selecting DoE that supports our investigation, we chose nested design because the prediction style (factor B) is nested within the learning approach (factor A). Note that the widely used cross design requires each factor to be applied equivalently across each level of the other factor, which does not apply to our study because the datasets for prediction styles M1-M3 are different for different learning approaches. That is, feature vectors have different values for change metrics and the predicted variables are different, as described in subsection 4.1.

Furthermore, our design is unbalanced because the number of observations (i.e., instances of machine learning experiments) for different combinations of factors' levels are different. Specifically, for within-release prediction styles (M1 and M2), the performance metrics (i.e., instances) were computed using the datasets of each of the 64 releases.

**Table 4**
The number of instances for the nested DoE

| Learning approach | | | | | |
|---|---|---|---|---|---|
| useAllPredictAll | | | usePrePredictPost | | |
| Prediction style | | | Prediction style | | |
| M1 | M2 | M3 | M1 | M2 | M3 |
| 63 | 64 | 52 | 58 | 59 | 46 |

However, for prediction across-releases (M3) the number of machine learning instances for each of the twelve projects (listed in Table 3) equals the number of its releases minus one, since it takes two releases to generate one prediction. In total, for the twelve projects considered in this paper, there were 52 instances for the across-releases predictions. It should be noted that some instances were excluded from the analysis because the classification performance was not reliable (e.g., all files were classified as fault-prone or not fault-prone). For example, for within-release with 10-fold cross validation (M1) prediction style, only one model was excluded for the *useAllPredictAll* learning approach (leading to 63 instances), while six models were excluded for *usePrePredictPost* (resulting in 58 instances). The numbers of instances for all combinations of the two learning approaches and three prediction styles are shown in Table 4[2].

Nested analysis of variance (ANOVA), also called a hierarchical ANOVA, is an extension of ANOVA for experiments where each group is divided into two or more subgroups. It tests if there is variation between groups, or within nested subgroups. Nested ANOVA (as ANOVA in general) is a parametric method which assumes population normality and variance homogeneity, and a balanced design is preferred. Since these assumptions do not hold in our case, we used the non-parametric alternative proposed in (Stavropoulos and Caroni, 2008) and used elsewhere (Zahalka et al., 2010). The Box-Adjusted Wald-type statistic neither assumes normality nor homogeneity of the data. It is also robust to unbalanced design. We used it to test the following null hypotheses, related to RQ1 and RQ2, respectively:

- $H_0^A$: There is no difference in the distributions of each performance metric (Recall, Precision, 1-FPR, F-Score and G-Score) between the two learning approaches, and
- $H_0^{B|A}$: For a given learning approach, there is no difference in the distribution of each performance metric among prediction styles M1, M2, and M3.

## 6. Results

In this section we report the results as they pertain to our research questions. Fig. 4 shows the box plots, and Fig. 5 depicts the means of all performance metrics for the two learning approaches and three prediction styles. The basic statistics (i.e., mean, median, standard deviation, and IQR) are provided in (Ahmad et al., 2022).

### 6.1. RQ1: Does the learning approach affect the classification performance of the software fault-proneness predictions?

As shown in Figs. 4 and 5, models based on *useAllPredictAll* had significantly higher Recall, Precision, F-Score and G-Score than the same models based on *usePrePredictPost*. However, 1-FPR was slightly lower when using *useAllPredictAll* than in the case of *usePrePredictPost*.

Table 5 shows the analysis of variance results for the DoE described in subsection 5.2. Since p-values were less than the significance level 0.05 for Recall, Precision, F-Score and G-Score, the null hypotheses $H_0^A$ were rejected in favor of the alternative hypotheses $H_a^A$ that there is a difference in the performance metrics distributions for each of these performance metrics. Furthermore, the learning approach factor contributed the most to the variance of the prediction performance, i.e., 76.4%, 72.99%, 67.57%, and 58.31% for Recall, Precision, F-Score, and G-Score, respectively. However, for the 1-FPR performance metric $H_0^A$ cannot be rejected, i.e., unlike the other performance metrics, 1-FPR was not affected by the learning approach.

Specifically, in prediction within-release using 10-fold cross validation (M1), *useAllPredictAll* compared to *usePrePredictPost* had 52.8% higher Recall, 38.8% higher Precision, 51.7% higher F-Score, and 22.4% higher G-Score. However, 1-FPR was 2.0% lower. In prediction within-release using 50/50 split (M2) the same pattern was observed, but the differences in performance were much higher. In particular, the *useAllPredictAll* learning approach had 77.6% higher Recall, 73.1% higher Precision, 79.5% higher F-Score, and 42.5% higher G-Score than *usePrePredictPost*. As in the case of 10-fold cross validation, 1-FPR was 2.2% lower. Similarly, in prediction across-release (M3), compared to *usePrePredictPost* learning approach, *useAllPredictAll* had a 111.1% higher Recall, 57.1% higher Precision, 97.3% higher F-Score, and 26.1% higher G-Score. However, it had 9.6% lower 1-FPR.

> Our findings thus provide a clearly affirmative answer to RQ1: the *useAllPredictAll* learning approach leads to better fault-proneness prediction performance than *usePrePredictPost*, both within-release and across-releases, for 4 of the 5 performance metrics.

### 6.2. RQ2: For a given learning approach, what is the difference in classification performance between using within-release and across-releases prediction styles?

To answer this research question, we again refer to Figs. 4 and 5. When the *useAllPredictAll* learning approach was used for within-release predictions, the models which used 10-fold cross validation (M1) slightly outperformed the corresponding models which used 50/50 split (M2) in terms of the mean Recall, Precision, F-Score, and G-Score, and had the same mean 1-FPR. Specifically, using 10-fold cross validation led to 2.1% higher Recall, 2.5% higher Precision, 1.7% higher F-Score, and 1.5% higher G-Score compared to when 50/50 split was used. Both within-release prediction styles (M1 and M2) significantly outperformed the across-releases prediction (M3) with respect to all performance metrics. Since the two within-release styles had similar performance, we restrict the comparison to using the 50/50 split. When compared to prediction across-releases, the performance within-release using 50/50 split had 23.9% higher Recall, 23.9% higher Precision, 15.3% higher 1-FPR, 33.8% higher F-Score, and 30.5% higher G-Score.

---
[2]Note that, following many related works (e.g., (Malhotra and Raje, 2015; Bowes et al., 2018)) that used the open-source datasets created by Jureczko and Spinellis (2010); Jureczko and Madeyski (2010), we treated the releases as independent instances for the statistical analysis.

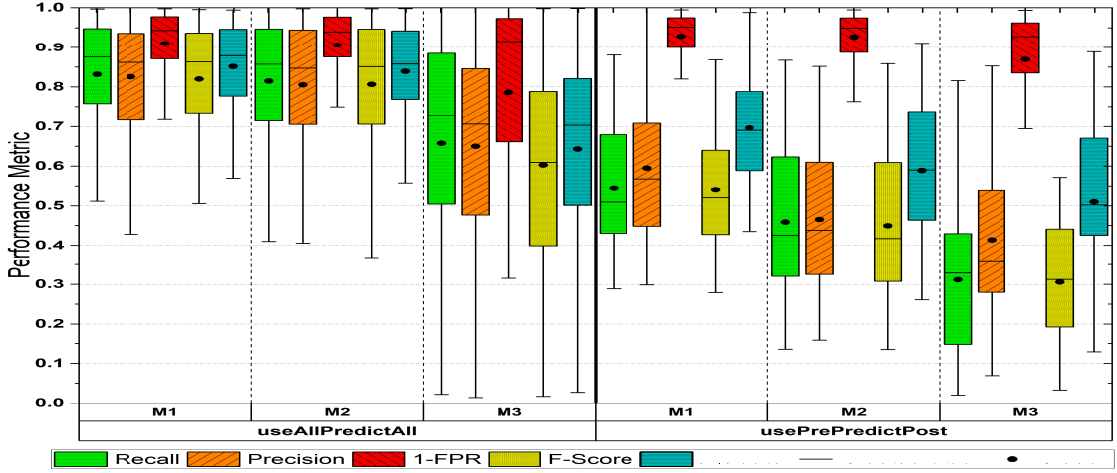

**Figure 4:** Performance metrics for the two learning approaches and three prediction styles

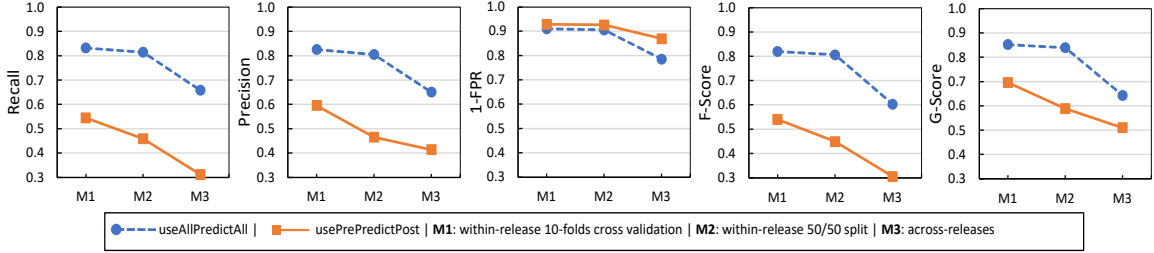

**Figure 5:** Means of all performance metrics for the two learning approaches and three prediction styles

**Table 5**
Analysis of variance results for learning approach ($A$) and nested prediction style ($B|A$)

| Performance Metric | Factor | Wald-type Box-Adjusted Rank Statistic | p-value | $H_0^A$ $H_0^{B|A}$ | Cont to Var % |
|---|---|---|---|---|---|
| Recall | $A$ | 39.65865 | 3.02E-10 | Rej | 76.40 |
|  | $B\|A$ | 7.86835 | 9.65E-02 | Not Rej | 23.60 |
| Precision | $A$ | 32.51898 | 1.18E-08 | Rej | 72.99 |
|  | $B\|A$ | 8.86537 | 6.46E-02 | Not Rej | 27.01 |
| 1-FPR | $A$ | 0.20742 | 6.49E-01 | Not Rej | 5.97 |
|  | $B\|A$ | 3.35987 | 5.00E-01 | Not Rej | 94.03 |
| F-Score | $A$ | 37.79243 | 7.87E-10 | Rej | 67.57 |
|  | $B\|A$ | 12.64919 | 1.31E-02 | Rej | 32.43 |
| G-Score | $A$ | 24.77202 | 6.45E-07 | Rej | 58.31 |
|  | $B\|A$ | 15.80304 | 3.30E-03 | Rej | 41.69 |

When *usePrePredictPost* was used for within-release predictions, the models which used 10-fold cross validation (M1) outperformed the models which used 50/50 split (M2) for all performance metrics. Specifically, using 10-fold cross validation within-release led to 18.7% higher Recall, 27.9% higher Precision, 0.2% higher 1-FPR, 20.3% higher F-Score, and 18.2% higher mean G-Score compared to the same models which used the 50/50 split. Consistent with the findings for the *useAllPredictAll* learning approach, when *usePrePredictPost* was used, both within-release prediction styles (M1 and M2) outperformed the across-releases prediction (M3) with respect to all performance metrics. In particular, when using within-release with 10-fold cross validation, compared to across-releases predictions the Recall, Precision, 1-FPR, F-Score, and G-Score were 74.8%, 43.8%, 6.8%, 76.9%, and 36.5% higher, respectively. Following the same pattern, using within-release with 50/50 split compared to across-releases predictions resulted in 47.2% higher Recall, 12.4% higher Precision, 6.6% higher 1-FPR, 47.0% higher mean F-Score, and 15.5% higher G-Score.

To further investigate the effect of the prediction style on the fault-proneness predictions for a given learning approach, we refer to the rows relevant to the hypothesis $H_0^{B|A}$ in Table 5. It appears that $H_0^{B|A}$ cannot be rejected for Recall,

Precision, and 1-FPR as the p-values were greater than the significance level 0.05. On the other hand, $H_0^{B|A}$ hypotheses were rejected for F-Score and G-Score in favor of the alternative hypotheses that, for a given learning approach, there is a difference in the performance due to prediction styles.

> Our findings give an affirmative answer to RQ2: the classification performance is significantly better when within-release prediction is used than when across-releases prediction is used, for both learning approaches.

Our answer to RQ2 is consistent with prior works for each given learning approach. However, our findings go a step further, showing that the contribution to the variance due to prediction style is smaller than the contribution to the variance due to the learning approach, as shown in Table 5. In other words, *the learning approach matters more than the prediction style*.

### 6.3. On reasons behind different performance

Our results for RQ1 showed that the prediction models which used *useAllPredictAll* had significantly better performance than the models which used *usePrePredictPost* in terms of Recall, Precision, F-Score and G-Score, but had similar or slightly worse 1-FPR. To better understand and explain the reasons behind the impact of the learning approach choice on software fault-proneness prediction performance, we conducted further investigation. Specifically, we took a closer look at the datasets used for each learning approach and hypothesized that the less imbalanced datasets used for the *useAllPredictAll* learning approach compared to the datasets used by the *usePrePredictPost* learning approach (see Figs. 2 and 3) cause *useAllPredictAll* to perform better.

In order to investigate if the different levels of class imbalance intrinsically present in the datasets used by different learning approaches cause the difference in performance, we re-evaluated our models using the Synthetic Minority Over-sampling Technique (SMOTE) (Chawla et al., 2002). SMOTE has been used in prior studies as a treatment for imbalance (e.g., (Agrawal and Menzies, 2018; Goseva-Popstojanova et al., 2019)). It works by selecting random data points from the minority class (i.e., bug-fixes when using *useAllPredictAll* and post-release bugs when using *usePrePredictPost*) and mimicking similar data points to over-represent the minority class, with the goal of enhancing the classification performance.

When using SMOTE, we oversampled the minority class (i.e., fault-prone files) in the training set to match the number of instances of the majority class. For example, when applying SMOTE for the *useAllPredictAll* learning approach with the Ant 1.3 release, the minority class was oversampled to increase the percentage of bug-fixes in the training sets from 44.7% to 50%. And when applying SMOTE for the *usePrePredictPost* learning approach, using oversampling the percentage of post-release bugs was increased from 6.5% to 50%. Note that the testing set was not modified, i.e., remained unbalanced, as in each original dataset. This eliminates over-fitting bias and yields more reliable results.

Applying SMOTE resulted in balanced datasets for both learning approaches, which we then used to re-run the experiments as described in Section 5. The box plots and means of all performance metrics when SMOTE was applied are shown in Figs. 6 and 7, respectively. In Table 6 we report the differences in performance metrics when SMOTE was applied compared to when SMOTE was not applied.

As can be seen from Figs. 6 and 7, and Table 6, applying SMOTE with the *useAllPredictAll* learning approach improved all performance metrics for within-release prediction styles (M1 and M2). Using SMOTE across-releases (M3) only slightly improved the Recall and 1-FPR, while it slightly decreased the Precision, F-Score and G-Score. Applying SMOTE with the *usePrePredictPost* learning approach led to a significant increase of Recall, Precision, F-Score, and G-Score in within-release prediction styles (M1 and M2), while the increase in across-releases predictions (M3) was minor. In all cases, applying SMOTE resulted in slightly lower 1-FPR.

In summary, using SMOTE for within-release predictions (M1 and M2) significantly improved all performance metrics expect 1-FPR, for both learning approaches. The improvement for *usePrePredictPost* was significantly higher than for *useAllPredictAll*. This is due to the fact that the datasets used with the former were significantly more imbalanced than the datasets used with the latter, leading to a much higher positive impact of the imbalance treatment.

Table 7 shows the results of the analysis of variance for our nested design when SMOTE was applied. The statistical results confirm the observations made based on the results shown in Figs. 6 and 7 – for all performance metrics we cannot reject the null hypothesis $H_0^A$ that there is no difference between the performance of the two learning approaches.

It should be noted that prior works have shown that treating the class imbalance improves the software fault-proneness prediction performance. Some of these works used *useAllPredictAll* (Wang and Yao, 2013; Malhotra and Jain, 2020; Goyal, 2022), while other used *usePrePredictPost* learning approach (Goseva-Popstojanova et al., 2019). What is new here is that we apply class imbalance treatment on both learning approaches and compare the corresponding improvements. Furthermore, we use imbalance treatment with a different goal than related works – to rigorously test our hypothesis that, for within-release predictions, the class imbalance is the hidden variable that explains the difference in prediction performance of the two learning approaches.

> For within-release predictions, our findings show that the worse performance of *usePrePredictPost* compared to the *useAllPredictAll* learning approach was due to the more pronounced class imbalance of the former. When addressed, using the imbalance treatment SMOTE, the performance difference between learning approaches was eliminated.

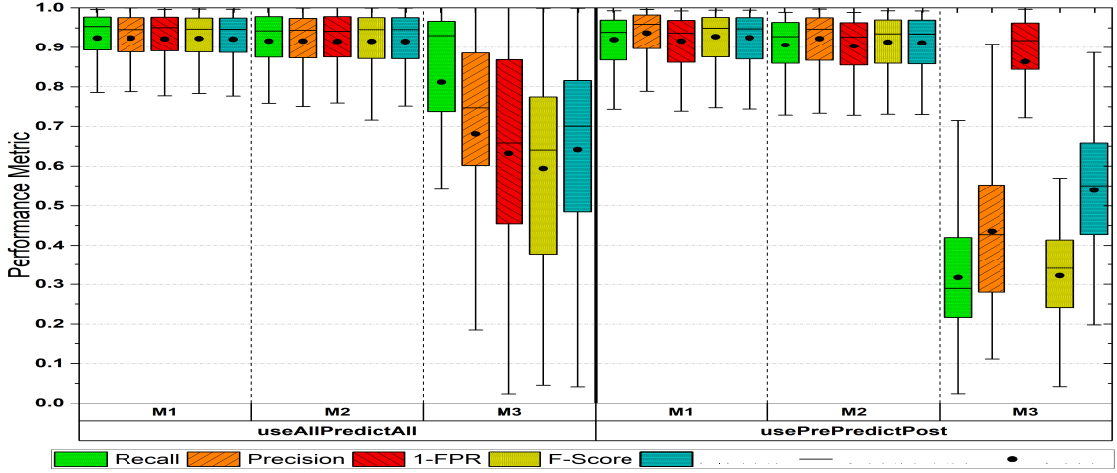

**Figure 6:** Performance metrics for the two learning approaches and three prediction styles when SMOTE was applied

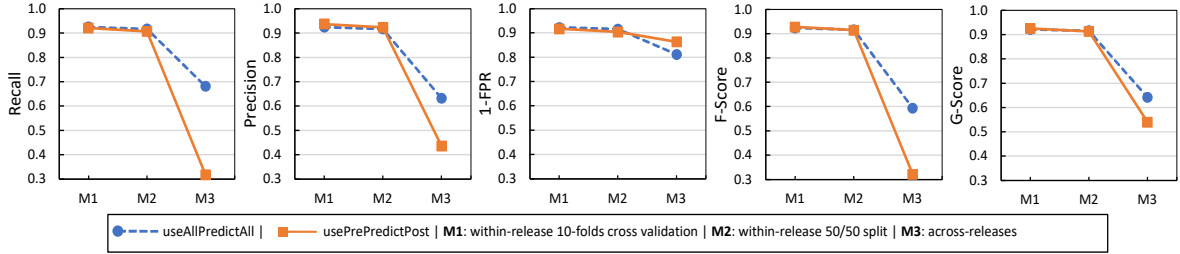

**Figure 7:** Means of performance metrics for the two learning approaches and three prediction styles when SMOTE was applied

**Table 6**
Differences in performance metrics when SMOTE was applied compared to when SMOTE was not applied

|  | useAllPredictAll | | | usePrePredictPost | | |
|---|---|---|---|---|---|---|
| Metric | M1 | M2 | M3 | M1 | M2 | M3 |
| Recall | 11.2% | 12.6% | 3.5% | 69.1% | 97.5% | 1.7% |
| Precision | 12.0% | 13.9% | -2.7% | 57.5% | 98.5% | 5.3% |
| 1-FPR | 1.4% | 1.2% | 3.3% | -1.3% | -2.5% | -0.7% |
| F-Score | 12.6% | 13.7% | -1.5% | 71.7% | 103.5% | 5.3% |
| G-Score | 8.2% | 9.1% | -0.2% | 33.0% | 54.9% | 5.8% |

**Table 7**
Analysis of variance results for learning approach ($A$) and nested prediction style ($B|A$) when SMOTE is applied

| Performance Metric | Factor | Wald-type Box-Adjusted Rank Statistic | p-value | $H_0^A$ $H_0^{B|A}$ | Cont to Var % |
|---|---|---|---|---|---|
| Recall | A | 2.864469 | 9.06E-02 | Not Rej | 10.55 |
|  | B \| A | 38.57545 | 8.52E-08 | Rej | 89.45 |
| Precision | A | 0.1863529 | 6.66E-01 | Not Rej | 0.76 |
|  | B \| A | 42.73856 | 1.17E-08 | Rej | 99.24 |
| 1-FPR | A | 0.7205299 | 3.95E-01 | Not Rej | 25.87 |
|  | B \| A | 2.089948 | 7.19E-01 | Not Rej | 74.13 |
| F-Score | A | 0.7934577 | 3.73E-01 | Not Rej | 3.15 |
|  | B \| A | 47.81298 | 1.03E-09 | Rej | 96.85 |
| G-Score | A | 0.518229 | 4.72E-01 | Not Rej | 2.08 |
|  | B \| A | 44.453 | 5.17E-09 | Rej | 97.92 |

Contrary to within-release predictions, using SMOTE for across-releases prediction (M3) had a minimal effect on prediction performance for both *useAllPredictAll* and *usePrePredictPost* learning approaches (see Table 6). As can be seen in Table 7, the null hypotheses $H_0^{B|A}$ were rejected for all performance metrics except 1-FPR in favor of the alternative hypotheses $H_a^{B|A}$ that (for a given learning approach) there is a difference in the performance metrics distributions due to the nested factor (i.e., prediction style). Moreover, when SMOTE was applied, the prediction style factor contributed most of the variance, i.e., 89.45%, 99.24%, 74.13%, 96.85%, and 97.92% in the case of Recall, Precision, 1-FPR, F-Score, and G-Score, respectively.

In general, prediction across-releases is more challenging than prediction within-release because it is harder for machine learning algorithms to learn effectively when the sources of training and testing data are different, which typically results in different distributions (Xu et al., 2018, 2019). It is an open research question whether different approaches for improving the performance of across-release predictions (Xu et al., 2018, 2019; Amasaki, 2020; Kabir et al., 2021) or transfer learning approaches (Ma et al., 2012; Nam et al., 2013, 2018)) would reduce the difference in performance when using different learning approaches.

# 7. Threats to validity

We discuss the threats to validity grouped into four categories: construct, internal, conclusion, and external validity.

**Construct validity** is concerned with whether we are measuring what we intend to measure. One threat to construct validity is related to the commits made to version control systems, which were used to extract the change metrics. As expected, data were missing from some commits. Commits with no revision numbers were given auto-generated unique ID numbers and were included in the study. On the other hand, commits with no lines of code reported as added or deleted were omitted from this study, which represented less than 3% of all commits made to all projects combined. Unlike change metrics, static code metrics represent a snapshot in time. For each release, the static code metrics were extracted from the latest available binaries on the release date, and they were the same for both learning approaches. The feature vectors used for the machine learning experiments in this paper integrate the static code metrics and change metrics as described in subsection 4.3. For the statistical tests, following the common practice in this area, we treated the releases as independent instances.

**Internal validity** is concerned with the effects of unknown impacts that might affect the independent and dependent variables. To ensure data quality, which is one of the major threats to internal validity, we extracted our own data from the online Apache repository. Note that we used a consistent set of static code and change metrics for all the projects and releases considered in this paper. Upon extracting the data, we implemented manual and automated sanity checks to verify the quality of the extracted metrics. In addition, for randomly selected files from each project included in this study, we manually verified and validated the values of the automatically extracted metrics.

**Conclusion validity** threats may impact the ability to draw correct and reliable conclusions. Some threats to conclusion validity are related to the way descriptive statistics are reported and statistical tests are being used. For descriptive statistics, we provided the box plots of performance metrics in this paper and reported their means, medians, standard deviations, and IQRs in the supplemental document online (Ahmad et al., 2022). For the inferential statistics, we used design of experiments approach and non-parametric tests suitable for our datasets. Another threat to conclusion validity is related to the data sample sizes. The work presented in this paper is based on 64 releases of 12 different open-source projects, which is comparable to or larger than the sample sizes used in related works in this area.

**External validity** is concerned with our ability to generalize our conclusions. The relatively large number of releases and projects that we studied provides some generalizability. However, this work was based only on open-source software written in Java and available on the Apache Projects web server. Therefore, we cannot claim that the results would be valid for software products implemented in other languages and/or from different application domains.

# 8. Implications, recommendations, and research directions

In this paper we seek to bring to the attention of the research and practitioners' communities the unreported effects of the learning approach on software fault-proneness predictions. This paper has presented evidence that current software prediction works do not adequately take into account the learning approach used. It also has shown how the effect of excluding this missing factor can inadvertently distort empirical observations.

Through empirical results and statistical tests presented in Section 6 of this paper, we show that the learning approach significantly affects the performance of software fault-proneness prediction. We also provide interpretation of our findings for research questions RQ1 and RQ2, and uncover the reason behind the difference in performance between the two learning approaches for within-release predictions. In this section, we discuss the implications of our findings, provide recommendations for designing, reporting, and comparing software fault-proneness prediction studies, and suggest some directions where further work is needed.

**Use the Learning Approach Appropriate for the Goal and Available Data.** Software fault-proneness prediction models using the *useAllPredictAll* approach predict fault-proneness for the entire period under consideration (e.g., release) and thus reflect mainly the *developers' viewpoint*. On the other hand, models using the *usePrePredictPost* learning approach predict software units (e.g., files) that are expected to be fault-prone post-release. It thus better reflects the *users' viewpoint* regarding the software's perceived reliability and may more readily prevent costly consequences from post-release failures.

**Address Class Imbalance.** Since datasets used for software fault-proneness prediction are usually imbalanced, treatment of imbalance (such as SMOTE) should be used. We have shown that it significantly improves the classification performance of within-release prediction style, for each learning approach. This confirms the results presented in several prior works (Wang and Yao, 2013; Agrawal and Menzies, 2018; Goseva-Popstojanova et al., 2019; Malhotra and Jain, 2020; Goyal, 2022). Furthermore, we have shown that the class imbalance is the hidden variable that causes the difference in the prediction performance of the two learning approaches and that addressing the imbalance results in similar prediction performance for both learning approaches.

**Always Specify the Learning Approach.** Our results have demonstrated that information regarding the learning approach is essential to understanding and explaining the performance of software fault-proneness prediction. However, such information is often missing or is merely implicit, in some cases requiring considerable effort to deduce. Therefore, research studies focused on software fault-proneness prediction, in addition to the existing reporting criteria (Hall et al., 2012), should explicitly specify the learning approach.

**Compare Apples-to-Apples.** To avoid comparing 'apples to oranges,' care should be taken when comparing the performance of software fault-proneness prediction models – when comparing to the related works' results, when conducting meta-analysis, and/or when reusing existing datasets that may have been created for different learning approaches. Our results may call into question conclusions advanced by some prior publications that failed to recognize the use of different learning approaches and consider their effect on the performance of the fault-proneness prediction.

We note that the abovementioned recommendations rely on existing approaches and techniques and therefore can be put into action right away.

Next, we present our viewpoint on the current-state-of-the art in software fault-proneness prediction and discuss some open issues for future research in this area.

**Current State-of-the-Art: software fault-proneness prediction models only consider spatial information** about software faults (i.e., their location) while disregarding their temporal information (i.e., when each fault was detected). Even though software fault-proneness prediction classifies software units (the "where"), the faults within fault-prone unit(s) were detected at different times. The complete lack of temporal information in the case of the *useAllPredictAll* learning approach leads (at least for some faults) to predicting the past from the future. This is not the case with the *usePrePredictPost* learning approach because it distinguishes between faults detected pre-release and post-release (i.e., it considers temporal information at a coarse-grained level) and predicts post-release fault-prone units.

**Future Research Direction: Account for timing information.** Incorporating temporal information about software faults into the fault-proneness prediction is essential to creating prediction models that are truer representations of the fault detection process. Doing so would address the fundamental drawback of the *useAllPredictAll* approach. Additionally, considering finer-grained temporal information (beyond the current distinction between pre-release and post-release faults in the *usePrePredictPost* approach) would help advance the software fault-proneness research area and increase its practical usefulness. Initial effort along these lines is presented in a recent study by Kabir et al. (2021) which focused on fault-proneness prediction for *useAllPredictAll* learning approach and across-releases prediction style, and conducted learning from data chunks that arrive in temporal order. In future work we plan to investigate incorporating both spatial and temporal information about software faults into software fault-proneness prediction.

Finally, **our findings provide some new insights for software developers**. The results of this study improve understanding and interpretability of software fault-proneness prediction, especially related to how the learning approach affects it. Better fault-proneness prediction provides more accurate information to developers, helping them prioritize development and testing tasks. Ultimately, improved prediction performance results in fewer bugs and better software quality. Consistently with prior works, our results showed that imbalance treatment (here using SMOTE) improved the Recall and Precision, resulting in higher F-Scores for both learning approaches when using the within-release prediction style. Further, we discovered that addressing the imbalance eliminated the inferior performance of the *usePrePredictPost* approach, which traditionally has had lower Recall than models built using the *useAllPredictAll* approach (e.g., (Zimmermann et al., 2007; Krishnan et al., 2013)). Reducing the number of post-release false negatives (due to higher Recall) is especially important for safety-critical systems where post-release failures can have catastrophic consequences. In addition, higher Precision leads to less wasted verification and validation efforts on those software units that likely are not fault-prone.

While this paper focuses on classification-based prediction within-projects (i.e., within-release and across-releases), we anticipate that the focus on the choice of the learning approach might yield new insights when exploring across-projects classification-based prediction or prediction of fault-counts in software units for different prediction styles.

## 9. Conclusion

This paper focuses on an aspect of the software fault-proneness prediction process – the learning approach – that was previously unexplored, except for one initial work.

First, we categorized related works by the learning approach each used, which clarifies and explains the varying performances of current software fault-proneness prediction works.

We then systematically and rigorously explored the impact of the learning approach on the performance of software fault-proneness prediction. Our study showed empirically that the choice of learning approach significantly affects the performance of software fault-proneness prediction models. As well, it produced new insights into how those effects occur. Our results showed that the *useAllPredictAll* learning approach resulted in significantly better performance than the *usePrePredictPost* learning approach. Furthermore, we uncovered that the class imbalance is the reason behind this finding for within-release predictions. Addressing the imbalance (in this paper by using SMOTE) significantly enhanced the within-release prediction performance for both learning approaches and eliminated the difference in their performance.

Finally, we described the implications of our findings; provided recommendations for designing, reporting, and comparing software fault-proneness prediction studies; and suggested some directions where further work is needed.

Going forward, we encourage fault-proneness prediction studies to always state explicitly which learning approach is being used. We also encourage increased awareness and further study of the consequences of the learning approach choice in practice.

## Acknowledgments

This research was supported in part by NASA's Software Assurance Research Program (SARP) under grant funded in FY21 and by National Science Foundation grants 1513717, 1545028 and 1900716.